\documentclass[letterpaper,twocolumn]{article}

\DeclareUnicodeCharacter{2212}{-}
\usepackage[T1]{fontenc}

\usepackage{geometry}
\usepackage{setspace}

\usepackage[style = chem-acs]{biblatex}
\usepackage{graphicx}

\usepackage{amsmath}
\usepackage{amssymb}
\usepackage{bm}

\usepackage{float}
\newfloat{scheme}{htbp}{los}
\floatname{scheme}{Scheme}
\floatname{chart}{Chart}
\newfloat{graph}{htbp}{loh}

\usepackage{chemformula} 
\usepackage[version = 4]{mhchem} 

\usepackage{authblk}
\author[1, 2]{Siyu Duan}
\affil[1]{Research Institute of Superconductor Electronics (RISE) \& Key Laboratory of Optoelectronic Devices and Systems with Extreme Performances of MOE,\par
School of Electronic Science and Engineering, Nanjing University, Nanjing 210023, China}
\affil[2]{Department of Physics, TU Dortmund University, 44227 Dortmund, Germany}
\author[1,*]{Jingbo Wu}
\author[1]{Xiaoqing Jia}
\author[1]{Hongkai Shi}
\author[1]{Haojie Li}
\author[2]{Anneke Reinold}
\author[1]{Huabing Wang}
\author[3]{Ilya M. Eremin}
\affil[3]{Institut für Theoretische Physik III, Ruhr-Universität Bochum, 44801 Bochum, Germany}
\author[2]{Götz S. Uhrig}
\author[1,*]{Biaobing Jin}
\author[2, 4,*]{Zhe Wang}
\affil[4]{Dortmund Center for Advanced Exploration of Dynamics Across Limits Using Spectroscopy (DAEDALUS), TU Dortmund University, 44227 Dortmund, Germany}

\title{Experimental signature of transient symmetry breaking in a cavity dirty-limit superconductor}
\date{*Email: jbwu@nju.edu.cn\\
               bbjin@nju.edu.cn\\
                 zhe.wang@tu-dortmund.de
}

\begin{document}

\twocolumn[
\maketitle

\begin{center}
\begin{minipage}{0.9\textwidth}

\begin{abstract}
  Transient states of matter far from equilibrium may exhibit physical properties beyond those allowed by the equilibrium-state crystalline symmetries.
We explore ultrafast and direct electronic excitations of transient states in a dirty-limit superconductor with time-resolved terahertz-pump terahertz-probe spectroscopy. By embedding the superconductor into a designed microcavity structure of metallic split-ring resonators, we can sensitively probe the pump field-induced transient variations of the superconducting states around the cavity's resonant frequency.
Our results show that the strong terahertz field can transiently modify the symmetries of the electronic subsystems via the injection of a picosecond supercurrent, leading to high-order nonlinear dynamical responses that are not compatible with the equilibrium-state symmetries, which evidences for transient symmetry breaking on the picosecond time scale.
\end{abstract}

\vspace{0.5em}

\end{minipage}
\end{center}

\vspace{1cm}
]

A paradigmatic classification of states of matter based on symmetries has long been established in condensed matter.
This paradigm also offers an elegant description of phase transitions accompanied by symmetry breaking.
Not only the equilibrium-state thermodynamic quantities governed by the symmetries are crucial for the description of a phase transition, but also the nonlinear dynamical responses are very sensitive to the changes of symmetry.
Recent experimental studies using time-resolved ultrafast spectroscopies and microscopies have revealed transient symmetry breaking in a variety of condensed matter systems (see e.g. \cite{Mitrano2016,Sie2019,Nova19,Li19,Yang2019,Sirica2022,Disa2023,
Ueda2023,Ilyas2024,Basini2024,Davies2024}), ranging from ferroelectrics \cite{Nova19,Li19,Basini2024}, (anti-)ferromagnetics \cite{Disa2023,Davies2024,Ilyas2024}, to topological matter \cite{Sie2019,Sirica2022}, and even superconductivity \cite{Mitrano2016,Yang2019}.
While the linear or nonlinear phononic coupling channels have often been exploited to transiently alter the concerned symmetries of the lattice or electronic responses, experimental endeavours towards a direct modification of the electronic subsystem's symmetry by ultrafast optical spectroscopy remain scarce.
 
Nonlinear optical effects in superconductors have been a subject of intensive theoretical studies, since they can contain important information complementary to the linear-response probes of the superconducting systems (see e.g. \cite{Cea16a,MoorTheorySHG2017,Wakatsuki17,Moore19,Silaev19,
Schwarz20,Tsuji20,Lorenzana20,Derendorf24,li2025terahertznonlinearresponsecuprate}).
For instance, in a single-band superconductor the Higgs amplitude fluctuations of the order parameter are not linearly coupled to electromagnetic waves \cite{PekkerVarma15,ShimanoTsuji20}, hence a third- or higher-order nonlinear-response probe is required to reveal the Higgs mode \cite{Cea16a,Silaev19,Schwarz20,Tsuji20};
In contrast to a linear electric susceptibility which is not forbidden in all crystal systems, a second-order nonlinear susceptibility vanishes identically under inversion symmetry, therefore second- or even-order nonlinear response probes sensitively the inversion-symmetry-breaking superconducting states \cite{Wakatsuki17,MoorTheorySHG2017,Moore19}.
While odd-order nonlinear responses were observed in inversion-symmetric superconductors (e.g. NbN \cite{Matsunaga12,MatsunagaHiggs2013,Matsunaga_NbN_Science,WangWang22,Katsumi24},  MgB$_2$ \cite{Kovalev21,Reinhoffer22,Katsumi25,Yuan2025}, and high-$T_c$ cuprates \cite{Chu20}),
equilibrium- or steady-state inversion symmetry breaking in the pseudogap region \cite{Zhao2017} or induced by DC supercurrent \cite{Nakamura2019,Nakamura2020} have been evidenced by finite second-order nonlinear susceptibilities.
For a superconductor in the clean limit, even transient inversion-symmetry breaking has been demonstrated \cite{Yang2019}, while for a dirty-limit superconductor it remains a pending problem whether its transient nonlinear optical response can exhibit any symmetry-breaking behavior on an ultrafast time scale.

Nonlinear transport in a dirty-limit superconductor is qualitatively different from that in the clean limit. In the clean limit, the superfluid density is high and the quasiparticle mean free path is long with the coherence length primarily determined by the pairing interaction.
Under these conditions, characteristic nonlinear optical response can arise from intrinsic band geometric effects or inversion-symmetry breaking, with minimal scattering to obscure the coherent response.
In contrast, a dirty-limit superconductor is characterized by strong impurity scattering, a substantially reduced mean free path (much shorter than the coherence length), and a modified electromagnetic response where the superfluid density is suppressed and the quasiparticle dynamics becomes diffusive.
Impurity scattering tends to average out anisotropic or symmetry-breaking effects, hence suppressing the corresponding nonlinear optical responses unless the underlying symmetry breaking is sufficiently robust against disorder.
In this work, we use a picosecond, linearly-polarized terahertz (THz) electric field to drive a dirty-limit superconductor NbN and another THz pulse to investigate the characteristic transient nonlinear response of the driven states. 
By designing and fabricating a microcavity structure based on NbN thin films, we are able to very sensitively detect the otherwise hardly resolved ultrafast even-order nonlinear effects, evidencing for THz field-induced transient inversion-symmetry breaking in the dirty limit.

The basic hypothesis underlying our time-resolved terahertz-pump terahertz-probe spectroscopic study is that the pump field-induced changes of the transmitted probe field $\delta E_{\text{probe}}(t_\text{pp})$ through the sample for a multicycle pump field $E_{\text{pump}}(f)$ with a central frequency $f$ can be described by 

{\footnotesize
\begin{equation}
\label{Eq_NL}
\begin{split}
\delta E_{\text{probe}}(t_\text{pp}) \sim 
[\chi^{(2)}E_{\text{pump}}(f,t_\text{pp}) + 
\chi^{(3)}E^2_{\text{pump}}(f,t_\text{pp})\\ 
+ \chi^{(4)}E^3_{\text{pump}}(f,t_\text{pp}) + \ldots] E_{\text{probe}},
\end{split}
\end{equation}
}
where $t_\text{pp}$ denotes pump-probe time delay, and $\chi^{(n)}$ with $n=2,3,4$ denotes the relevant tensor components in the $n^\text{th}$-order nonlinear susceptibilities of the cavity-superconductor device.
The $1f$, $2f$, and $3f$ components in $\delta E_{\text{probe}}$ are governed by the second-, third-, and fourth-order nonlinear susceptibilities $\chi^{(2)}$, $\chi^{(3)}$, and $\chi^{(4)}$, respectively.
In this phenomenological description of the experimental data, we have taken the following facts into consideration: Firstly, the detected probe field is perpendicular to the pump-pulse polarization; Secondly, the pump field is stronger than the probe pulse by more than one of order of magnitude; Thirdly, the observed higher-order harmonic oscillations are weaker than the lower-order ones.
Due to the presence of inversion symmetry in NbN, the even-order susceptibilities should vanish, while the odd-order ones are allowed \cite{Tsuji20}.
Therefore, an observation of a strong coherent $1f$ and $3f$ modulations, corresponding to the even-order nonlinear responses governed by $\chi^{(2)}$ and $\chi^{(4)}$ that are forbidden at thermal equilibrium, will be evidence of transient inversion symmetry breaking.

\begin{figure}[!t]
\centering
\includegraphics[width=\columnwidth]{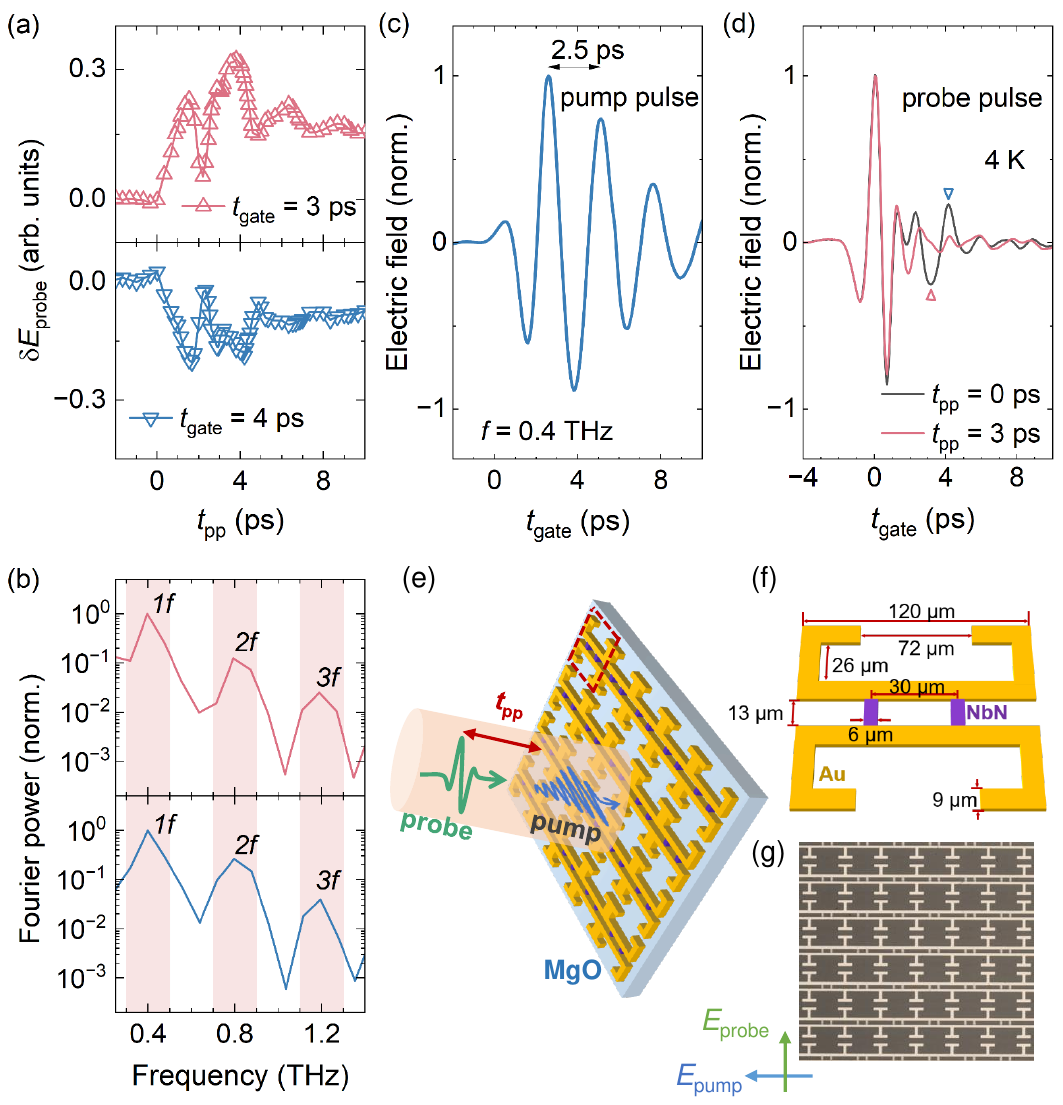}
\caption{\label{Fig_0.4THz4K}
(\textbf{a}) 0.4~THz pump-induced changes $\delta E_{\text{probe}}$ of the transmitted probe electric field as a function of pump-probe delay $t_{\text{pp}}$ at the time $t_{\text{gate}}=3$ and 4~ps corresponding to the terahertz probe pulse as marked by the triangles in (\textbf{d}).
(\textbf{b}) The frequency-domain spectra corresponding to the traces in (\textbf{a}) exhibit maxima at the fundamental, second, and third harmonics of the pump pulse, as marked by $1f$, $2f$, and $3f$, respectively,  which correspond to the second-, third-, and fourth-order nonlinear susceptibilities $\chi^{(2)}$, $\chi^{(3)}$, and $\chi^{(4)}$ in Eq.~(\ref{Eq_NL}).
(\textbf{c}) Time-domain profile of the 0.4~THz pump pulse.
(\textbf{d}) Time-domain traces of the transmitted probe pulses measured at 4~K for pump-probe time-delays $t_{\text{pp}}=0$ and 3~ps.
(\textbf{e}) Schematic illustration of the pump-probe experiment with normal incidence.
(\textbf{f}) Design and dimensions of one cavity unit based on gold microstructure and  12~nm thick NbN superconductor thin film grown on $\langle100\rangle$-oriented 0.5~mm thick MgO substrate.
(\textbf{g}) Microscope image of the structured sample with the arrows indicating the pump and probe field polarizations. 
}
\end{figure}

Our time-resolved THz pump-THz probe spectroscopy was built based on a femtosecond laser with 800~nm central wavelength and 1~kHz repetition rate. 
Terahertz probe pulses were generated using a ZnTe crystal (see Supporting Information), while strong THz pump pulses were generated in a LiNbO$_3$ crystal via tilted-pulse-front optical rectification. 
Narrowband pump pulses were obtained by using THz bandpass filters.
Polarizations of the pump and probe fields were tuned with wire-grid polarizers to be perpendicular to each other, and aligned to the high-symmetry directions as indicated in Figure ~\ref{Fig_0.4THz4K}(g).
The transmitted field of chopper-modulated probe pulses in normal incidence to the sample was detected by free-space electro-optic sampling in another ZnTe crystal \cite{Zhang96} with lock-in amplifier.
The low-temperature measurements were carried out in a helium-flow optical cryostat.

For a bare, dirty-limit NbN film under drive of a THz electric field, an even-order nonlinear response was not resolvable \cite{Yang2019}.
This, on the one hand, indicates (if not vanishing) a very small transient inversion-symmetry breaking effect for a dirty-limit superconductor due to strong scattering. 
On the other hand, the sensitivity of a THz probe pulse is substantially reduced due to the very low transmission of the THz pulse through a dirty-limit superconductor \cite{DemsarNbNPRL2011}.
To improve the signal-to-noise ratio we design a THz sensitive microcavity structure by embedding NbN microbridges between metallic split-ring resonators [see Figure~\ref{Fig_0.4THz4K}(f) for its dimensions].
The microcavity structure is designed according to the following considerations.
Firstly, the microstructure itself is highly symmetric.
Secondly, the resonant frequency of the metallic split-ring resonators is 0.5~THz, well within the frequency range of our probe pulse.
Thirdly, the NbN microbridges control the coupling between two metallic split-ring resonators, allowing a tuning of the linear transmission at the resonant frequency of 0.5~THz by a factor of $\approx 5.5$ (see Supporting Information).
In comparison with a bare NbN film, such a microcavity device not only enhances transmission in the concerned frequency range, but it also converts transient variations of the NbN superconducting states into pronounced spectral changes of the THz probe pulse.

For an $f=0.4$~THz pump pulse [Figure~\ref{Fig_0.4THz4K}(c)] with a peak field of $\sim$~6.1~kV/cm, the pump field-induced changes $\delta E_{\text{probe}}$ of the transmitted probe electric field are presented in Figure~\ref{Fig_0.4THz4K}(a) as a function of pump-probe delay $t_\text{pp}$ measured below $T_c$ at 4~K [see Figure~\ref{Fig_0.4THz4K}(e) for an illustration]. The corresponding Fourier power spectra are displayed in Figure~\ref{Fig_0.4THz4K}(b) for the representative gate times $t_\text{gate}= 3$ and 4~ps of the probe pulse [see Figure~\ref{Fig_0.4THz4K}(d)], where the gate times $t_\text{gate}$ correspond to the time delays of a sampling laser pulse with respect to the THz probe pulse in the electro-optic sampling detection of the THz probe electric field~\cite{Zhang96}.
Time-domain traces of the probe pulses are shown in Figure~\ref{Fig_0.4THz4K}(d) (see also Supporting Information).

For both $t_\text{gate}= 3$ and 4~ps the pump-induced signals $\delta E_{\text{probe}}$ exhibit clear oscillation behavior with the change of the pump-probe delay $t_\text{pp}$. In the Fourier spectra we can clearly see peaks at 0.4, 0.8, and 1.2~THz, corresponding to one, two, and three times the pump frequency, as marked by $1f$, $2f$, and $3f$, respectively, in Figure~\ref{Fig_0.4THz4K}(b).
While the $2f$ oscillation corresponds to a third-order nonlinear response which is allowed by symmetry, our observation of the $1f$ and $3f$ signals is the most important finding of the present experimental study.
The $1f$ and $3f$ modulations of the probe pulse correspond to even-order nonlinearity (see Eq.~\ref{Eq_NL}), which is not allowed by the inversion symmetry at equilibrium state and has not been observed before in a superconductor in the dirty limit.

\begin{figure*}[t]
\centering
\includegraphics[width=0.88\textwidth]{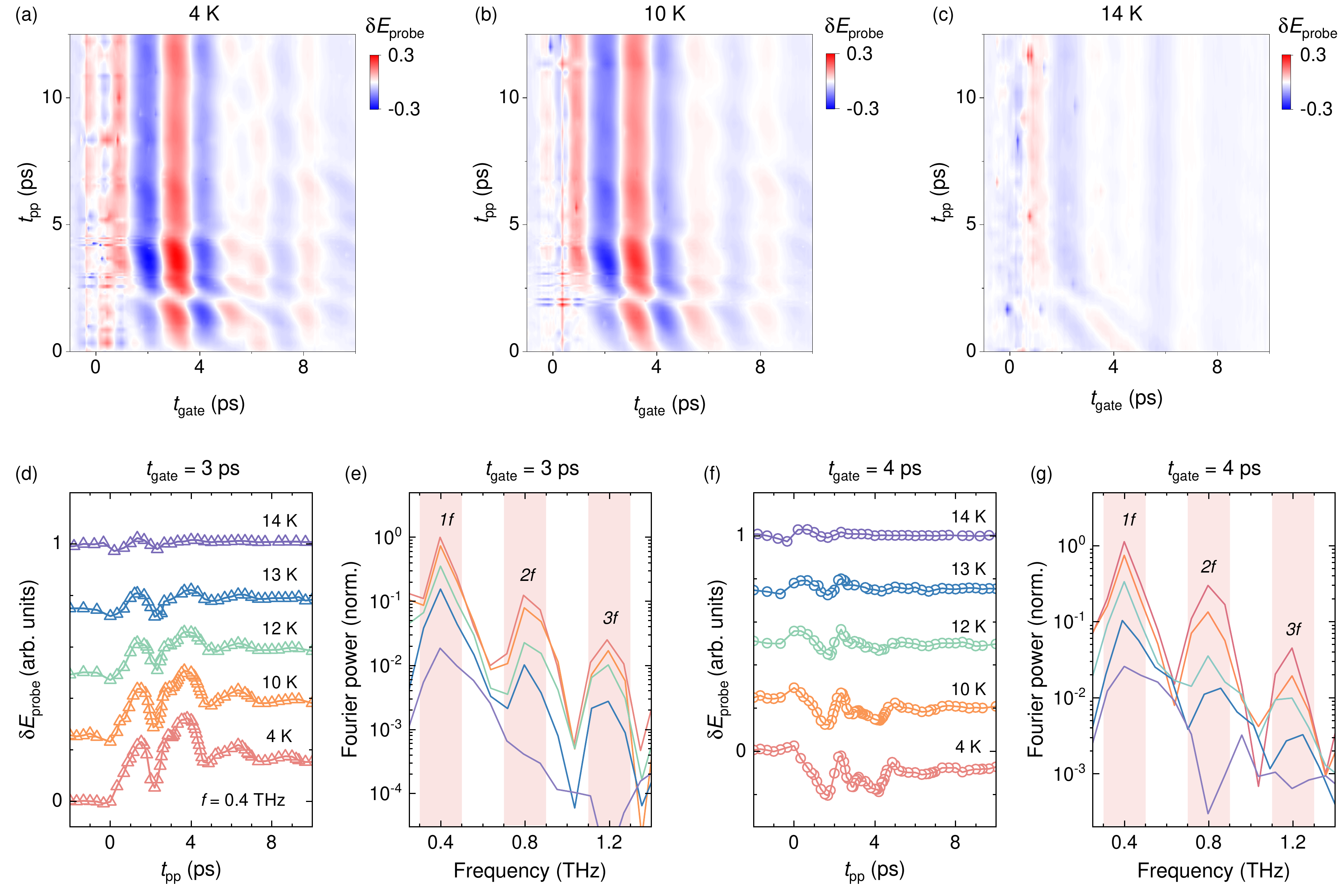}
\caption{\label{Fig_Tdep}
Contour plot of $f=0.4$~THz pump pulse induced temporal changes $\delta E_{\text{probe}}$ of the transmitted probe electric field as a function of pump-probe delay $t_{\text{pp}}$ measured at (\textbf{a}) 4~K, (\textbf{b}) 10~K, and (\textbf{c}) 14~K.
(\textbf{d}) Temperature dependence of the pump-field induced transient changes $\delta E_{\text{probe}}$ at $t_\text{gate}=3$~ps [Figure~\ref{Fig_0.4THz4K}(d)] and (\textbf{e}) the corresponding Fourier power spectra measured at various temperatures above and below $T_c$.
(\textbf{f})Temperature dependence of the pump-field induced transient changes $\delta E_{\text{probe}}$ at $t_\text{gate}=4$~ps [Figure~\ref{Fig_0.4THz4K}(d)] and (\textbf{g}) the corresponding Fourier power spectra measured at various temperatures above and below $T_c = 13.5$~K.
}
\end{figure*}

In order to clarify the relation between the transient symmetry breaking and the superconductivity, we measure the pump-probe signals at different temperatures below and above the superconducting transition at $T_c=13.5$~K.
Contour plots of the obtained pump-probe signals $\delta E_{\text{probe}}$ are presented in Figure~\ref{Fig_Tdep}(a)–\ref{Fig_Tdep}(c) for 4, 10, and 14~K, respectively.
At 4~K the pump-induced oscillations are observed not only at $t_\text{gate}= 3$ and 4~ps [Figure~\ref{Fig_0.4THz4K}(a)], but also for other $t_\text{gate}$'s where the probe fields are well resolved [Figure~\ref{Fig_0.4THz4K}(d)]. 
With increasing temperature below $T_c$, the oscillation behavior is still clearly visible [see Figure~\ref{Fig_Tdep}(b) for 10~K data].
However, already at 14~K, which is just above the superconducting phase transition, the pump-induced oscillation is nearly indiscernible in the contour plot. Therefore, the substantial nonlinear responses appear in the superconducting state.

To provide a detailed inspection of the temperature dependent behaviors, we present the pump-induced modulations for $t_\text{gate}=3$ and 4~ps measured at different temperatures crossing $T_c$ in Figure~\ref{Fig_Tdep}(d) and~\ref{Fig_Tdep}(f), respectively, and the corresponding Fourier power spectra in Figure~\ref{Fig_Tdep}(e) and~\ref{Fig_Tdep}(g).
Above $T_c$ at 14~K, the pump-induced changes are very small and merely visible [Figure~\ref{Fig_Tdep}(d) and~\ref{Fig_Tdep}(f)], which exhibits a very weak oscillation corresponding to the pump frequency [i.e. $1f = 0.4$~THz in Figure~\ref{Fig_Tdep}(e) and Figure~\ref{Fig_Tdep}(g)].
With entering the superconducting phase, the pump-induced signals are getting stronger and exhibit richer oscillation features [see 13~K curves in Figure~\ref{Fig_Tdep}(d) and Figure~\ref{Fig_Tdep}(f)].
These features are enhanced significantly and monotonically with decreasing temperature in the superconducting phase. As shown in the Fourier power spectra in Figure~\ref{Fig_Tdep}(e) and~\ref{Fig_Tdep}(g), whereas above $T_c$ only the fundamental frequency component is present, the pump-driven oscillation behavior exhibits also the $2f$ and the $3f$ components right below $T_c$.
In comparison with the 14~K spectra, the nonlinear responses at 4~K are enhanced by nearly two orders of magnitude, clearly evidencing the dominant role of superconductivity.
The enhancement of the nonlinear signal by 1--2 orders of magnitude cannot simply be due to the suppression of thermal fluctuations with decreasing temperature, because the temperature change is small. The observed weak \textit{1f} signal above $T_c$ cannot be assigned as just leakage of the pump field, since the pump field was not mechanically modulated by a chopper and essentially not resolvable in our detection scheme based on lock-in amplification (see Figure~S5 in Supporting Information).
We would speculate that very close to the superconducting transition temperature, the normal-state current due to the strong terahertz driving field could possibly induce weak transient symmetry breaking, which is captured by our probe field.
The observed strong temperature dependence of the \textit{1f} signal also rules out that this signal is due to leaked pump field reaching the electro-optic sampling crystal.

\begin{figure}[t]
\centering
\includegraphics[width=\columnwidth]{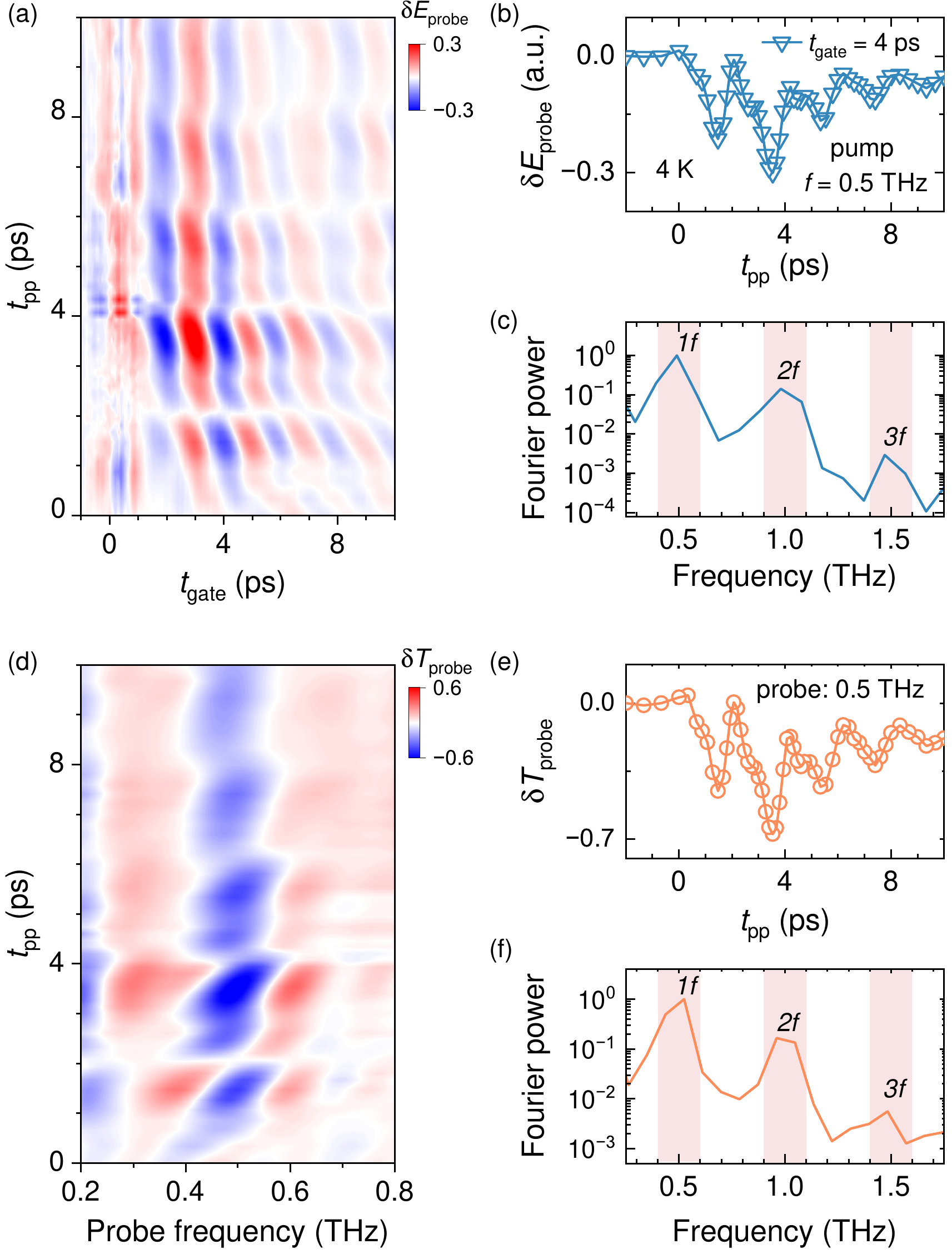}
\caption{\label{Fig_0.5THz}
(\textbf{a}) Contour plot of $f=0.5$~THz pump-induced temporal changes $\delta E_{\text{probe}}$ of the transmitted probe electric field as a function of pump-probe delay $t_{\text{pp}}$ measured at 4~K. 
(\textbf{b}) The corresponding nonlinear temporal modulation at $t_\text{gate}=4$~ps and (\textbf{c}) its Fourier power spectrum.
(\textbf{d}) Contour plot of $f=0.5$~THz pump-induced spectral changes of the transmission $\delta T_{\text{probe}}$ as a function of pump-probe delay $t_{\text{pp}}$ measured at 4~K. 
(\textbf{e}) The corresponding nonlinear spectral modulation at $0.5$~THz and (\textbf{f}) its Fourier power spectrum.
}
\end{figure}
 
The nonlinear modulations of the THz probe field by the excited cavity superconductor are not only observed in the time domain, but also for the frequency domain.
For a pump pulse of $f=0.5$~THz with a peak field of $\sim$~5.0~kV/cm, the temporal and spectral modulations are observed and presented by the pump-induced changes $\delta E_{\text{probe}}$ in Figure~\ref{Fig_0.5THz}(a) and $\delta T_{\text{probe}}$ in Figure~\ref{Fig_0.5THz}(d), respectively, where $T_{\text{probe}}$ denotes the transmission of the probe pulse at different probe frequencies.
A representative curve of the temporal modulation $\delta E_{\text{probe}}$ corresponding to $t_\text{gate}= 4$~ps is shown in Figure~\ref{Fig_0.5THz}(b).
The pump-field driven oscillation is clearly observed in the time domain, which contains not only the $2f$, but also the $1f$ and $3f$ components, as shown in the Fourier power spectrum in Figure~\ref{Fig_0.5THz}(c).

The pump-field driven spectral modulations are represented by the changes of transmission $\delta T_{\text{probe}}$ in the frequency domain as a function of pump-probe delay [Figure~\ref{Fig_0.5THz}(d)].
One can see evident spectral modulations in the available frequency range, indicating a coherent nonlinear response of the cavity superconductor driven by the THz electromagnetic wave.
As shown in Figure~\ref{Fig_0.5THz}(e) for a probe frequency of 0.5~THz, the transmission of the probe pulse exhibits very evident oscillations on the pump-probe delay $t_\text{pp}$.
The corresponding power spectrum displays also the $1f$, $2f$, and $3f$ components, as shown in Figure~\ref{Fig_0.5THz}(f), confirming the observation of the coherent second-, third-, and fourth-order nonlinear responses [see Eq.~(\ref{Eq_NL})].
As will be further discussed below, the capability to resolve the large modulation due to the transient supercurrent induced symmetry breaking is attributed to the substantially enhanced sensitive detection around 0.5~THz -- the resonant frequency of our designed microcavity structure (see Figure S2 in Supporting Information).

While third-order nonlinear responses have been observed in previous experimental studies of NbN \cite{Matsunaga_NbN_Science,Yang2019}, a second-order nonlinear response emerge only when the inversion symmetry is broken, for example, in a steady state under a DC supercurrent injection \cite{Nakamura2020}.
Therefore, the observed coherent $1f$ and $3f$ temporal and spectral modulations in our experiment are the signature of transient symmetry breaking induced by the picosecond multicycle THz electric field in the dirty-limit superconductor.
The results indicate that although the transport of the injected picosecond supercurrent experiences strong scattering, at least on the picosecond time scale a symmetry breaking effect is detectable, especially with the help of the THz resonator cavity structure. 

We note that the observed $1f$ oscillation in the superconducting state cannot be governed by extrinsic effects such as leakage of the pump field or asymmetry of the cavity structure, since the pump field is perpendicular to the probe field direction and the cavity structure is highly symmetric [Figure~\ref{Fig_0.4THz4K}(f),~\ref{Fig_0.4THz4K}(g)]. 
This is also confirmed by the experimental observation that the $1f$ oscillation reduces significantly from 4 K to 14 K by nearly two orders of magnitude [see Figure~\ref{Fig_Tdep}(e),~\ref{Fig_Tdep}(g)], following the temperature dependent evolution of the superconducting order parameter, whereas a linear term proportional to the pump field $E_{\text{pump}}$ should increase with increasing temperature towards the normal state due to increased transmission. 
Moreover, the $1f$ oscillation should not be dominated by the high-order nonlinear terms in $E_\text{probe}$, such as $\chi^{(3)}E_{\text{pump}}E^2_{\text{probe}}$.
Otherwise, the $2f$ oscillation governed by $\chi^{(3)}E^2_{\text{pump}}E_{\text{probe}}$ would be stronger than the $1f$ oscillation by at least one order of magnitude due to $E_{\text{pump}} \gg E_{\text{probe}}$, in contradiction to the experimental observations [see Figure~\ref{Fig_0.4THz4K}(b), Figure~\ref{Fig_Tdep}(e),~\ref{Fig_Tdep}(g), and Figure~\ref{Fig_0.5THz}(c),~\ref{Fig_0.5THz}(f)].
As shown in Figure~\ref{Fig_Tdep}(e) and~\ref{Fig_Tdep}(g), also the third- and fourth-order nonlinear responses ($2f$ and $3f$ oscillations) decrease drastically with increasing temperature, becoming not resolvable above $T_c$.
Therefore, the observed transient symmetry breaking effects are inherent characteristics of the superconducting state.
We would also note that the theoretical analysis is based on the assumption of perturbative nonlinear response as expressed in Eq.~(\ref{Eq_NL}), while the system under the drive may be in a weakly nonperturbative regime. It remains a very interesting problem that how the nonlinear response may evolve under a stronger terahertz field drive (see e.g. Ref.~\cite{Duan26a}).

\begin{figure}[t]
\centering
\includegraphics[width=\columnwidth]{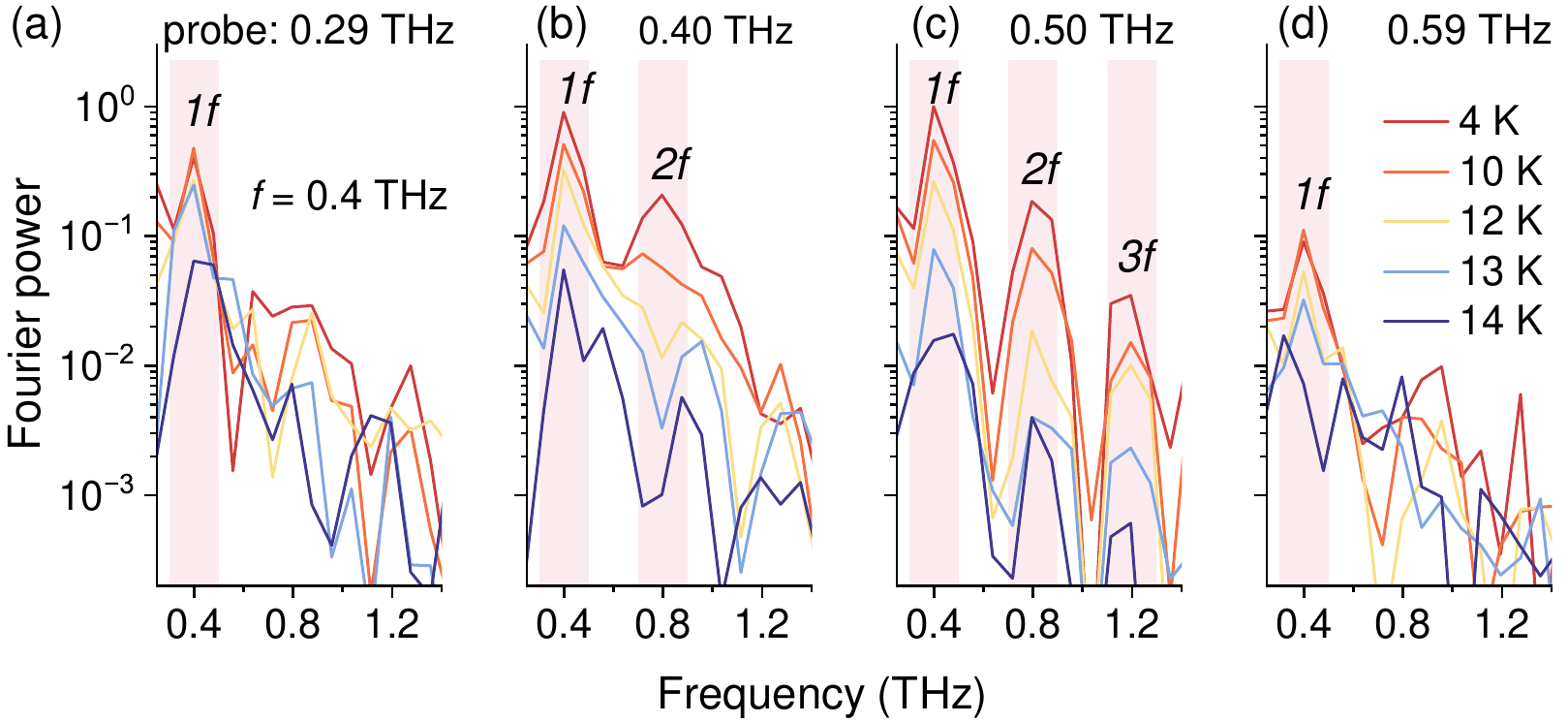}
\caption{\label{Fig_ProbeFrequencies}
Fourier power spectra of pump-induced modulation corresponding to the probe frequencies of (\textbf{a}) 0.29, (\textbf{b}) 0.40, (\textbf{c}) 0.50, and (\textbf{d}) 0.59~THz, respectively, for the pump frequency of $f=0.4$~THz, measured at various temperatures, indicating enhanced spectral detection around 0.5~THz corresponding to the resonant frequency of the microcavity structure.
}
\end{figure}

Since our designed metallic planar resonators have a resonant linear electromagnetic response at 0.5~THz (see Figure S2 in Supporting Information), we expect that the microcavity device is mostly responsive to the nonlinear modulation of THz probe fields around this frequency.
For various probe frequencies, the pump-induced modulations of their transmissions are presented in Figure~\ref{Fig_ProbeFrequencies}(a)–\ref{Fig_ProbeFrequencies}(d) as by the corresponding Fourier power spectra measured at different temperatures for a pump frequency of $f=0.4$~THz.
At the resonance probe frequency of 0.5~THz [Figure~\ref{Fig_ProbeFrequencies}(c)], the cavity device at the lowest temperature is sensitive to all modulations at $1f$, $2f$, and $3f$, similar to the observation for a 0.5~THz pump [see Figure~\ref{Fig_0.5THz}(f)].
With increasing temperature, the modulation intensity decreases, more rapidly for the higher-order responses.
Deviating from the resonance frequency, the probes are less sensitive and the observed pump-induced modulations are reducing.
For instance, for the probe frequency of 0.40~THz [Figure~\ref{Fig_ProbeFrequencies}(b)], the $3f$ modulations become invisible even at the lowest temperature, while the $2f$ modulation is visible only below 10~K.
At probe frequencies of 0.29 and 0.59~THz, only the lowest-order $1f$ modulation remains resolvable at low temperatures [see Figure~\ref{Fig_ProbeFrequencies}(a) and~\ref{Fig_ProbeFrequencies}(d).
These behaviours highlight the cavity-enhanced sensitive detection of the transient symmetry breaking states in the relevant THz spectral range, as well as the functionality of the designable cavity-superconductor metamaterial. 

Our work demonstrates that a cavity-engineered superconductor microstructure enables a very sensitive detection of THz-field induced transient symmetry breaking states.
The results of our studies not only show that transient symmetry breaking can be induced by picosecond supercurrent even in a dirty-limit superconductor with strong scattering, but also offer an all-optical, contact-free means of probing transient symmetry-breaking superconducting states.
Our findings on the nonlinear terahertz responses enrich the possibilities of engineering superconductor-based symmetry-breaking metamaterials for ultrafast applications \cite{Schalch19,Pettine2023,Duan2024,Srivastava2025,Aigner2025}.

\section{Supporting Information Available:}
Additional details of device fabrication, probe- and pump-pulse traces, linear transmission spectra of the cavity superconductor device, simulated electric field distributions, and sketch of the experimental setup, including Figs. S1--S6.

\section*{Acknowledgements}

We thank Zhiyuan Sun for valuable discussions, and acknowledge support by the European Research Council (ERC) under the Horizon 2020 research and innovation programme, Grant Agreement No. 950560 (DynaQuanta), and by the National Nature Science Foundation of China (62288101, 62331015, 62027807), Jiangsu Science Fund for Distinguished Young Scholars (BK20240062).
S.D. was supported by the Walter Benjamin Programme of the German Research Foundation (DFG) via the Project Number 557566396.

\printbibliography

\clearpage

\end{document}